\documentclass[traditabstract]{aa} 
\usepackage{graphicx}
\usepackage{txfonts}
\usepackage{natbib}
\usepackage{longtable}
\usepackage{amsmath}

\bibpunct[]{(}{)}{;}{a}{}{,}

\begin{document}

\title{Rotational spectroscopy of CH$_3$OD with a reanalysis of CH$_3$OD toward IRAS 16293$-$2422\thanks{Electronic supplementary material for this work can be found at https://doi.org/10.5281/zenodo.11460242}}

\author{
           V.~V. Ilyushin\inst{1}
           \and
           H.~S.~P. M{\"u}ller\inst{2}
           \and
           M.~N. Drozdovskaya\inst{3}
           \and
           J.~K. J{\o}rgensen\inst{4}
           \and
           S. Bauerecker\inst{5}
           \and
           C. Maul\inst{5}
           \and
           R. Porohovoi\inst{1}
           \and
           E.~A. Alekseev\inst{1,6}
           \and
           O. Dorovskaya\inst{1}
           \and
           O. Zakharenko\inst{2}
           \and
           F. Lewen\inst{2}
           \and
           S. Schlemmer\inst{2}
           \and
           L.-H. Xu\inst{7}
           \and
           R.~M. Lees\inst{7}
           }

   \institute{Institute of Radio Astronomy of NASU, Mystetstv 4, 61002 Kharkiv, Ukraine\\
              \email{ilyushin@rian.kharkov.ua}
              \and
              Astrophysik/I.~Physikalisches Institut, Universit{\"a}t zu K{\"o}ln,
              Z{\"u}lpicher Str. 77, 50937 K{\"o}ln, Germany\\
              \email{hspm@ph1.uni-koeln.de}
              \and
              Physikalisch-Meteorologisches Observatorium Davos und Weltstrahlungszentrum (PMOD/WRC), Dorfstrasse 33, CH-7260, Davos Dorf, Switzerland              
              \and
              Niels Bohr Institute, University of Copenhagen, {\O}ster Voldgade 5$-$7, 1350 Copenhagen K, Denmark 
              \and
              Institut f{\"u}r Physikalische und Theoretische Chemie, 
              Technische Universit{\"a}t Braunschweig, Gau{\ss}str. 17, 38106 Braunschweig, Germany
              \and
              Univ. Lille, CNRS, UMR 8523 - PhLAM - Physique des Lasers Atomes et Mol{\'e}cules, F-59000 Lille, France
              \and
              Department of Physics, University of New Brunswick, Saint John, NB E2L 4L5, Canada
              }

   \date{Received 09 Mar 2024 / Accepted 27 May 2024}
 
\abstract

\abstract{
We have started a measurement campaign of numerous methanol isotopologs in low-lying torsional states 
in order to provide extensive line lists for radio astronomical observations from an adequate 
spectroscopic model and to investigate how the intricate vibration-torsion-rotation interactions 
manifest themselves in the spectra of different isotopic species. 
After CD$_3$OH and CD$_3$OD, we turn our focus to CH$_3$OD, which is an important species 
 for studying deuteration in prestellar cores and envelopes that enshroud protostars. 
Notably, deuteration is frequently viewed as a diagnostic tool for star formation. The measurements used in this study were obtained in two spectroscopic laboratories and cover large fractions 
of the 34~GHz$-$1.35~THz range. As done in previous studies, we employed a torsion-rotation Hamiltonian model for our analysis that is based on the rho-axis method. 
The resulting model describes the ground and first excited torsional states of CH$_3$OD well 
up to quantum numbers $J \leqslant 51$ and $K_a \leqslant 18$. We derived a line list 
for radio astronomical observations from this model that is accurate up to at least 1.35~THz 
and should be sufficient for all types of radio astronomical searches for this methanol isotopolog 
in these two lowest torsional states. This line list was applied to a reinvestigation of CH$_3$OD 
in data from the Protostellar Interferometric Line Survey of IRAS 16293$-$2422 obtained with 
the Atacama Large Millimeter/submillimeter Array. The new accurately determined value for the column density of CH$_3$OD implies that the deuteration in methanol differs in its two functional groups by a factor of $\sim$7.5. 
}


\authorrunning{V.~V. Ilyushin et al.}
\titlerunning{Rotational spectroscopy of CH$_3$OD}

\maketitle
\hyphenation{For-schungs-ge-mein-schaft}

\section{Introduction}
\label{intro}

The singly deuterated methanol isotopolog CH$_3$OD was detected unambiguously by \citet{det_CH3OD_1988}, 
about twenty years after the detection of CH$_3$OH \citep{det_CH3OH_1970}. Since then, CH$_3$OD has become an important 
diagnostic tool for the degree of deuteration in star-forming regions \citep{D-MeOH_various_2011,D-MeOH_Orion-KL_2013,deuteration_NGC6334_2018,deuteration_16293_2018,deuterated_CH3OH_2019,methanol-deuteration_Orion-KL_2022,methanol-deuteration_B335_2022}. The degree of deuteration in turn has been considered an indicator of the conditions of star formation \citep{deuteration_2005,deuteration_2007,deuteration_2011,deuteration_2018} and has even been used to estimate the age of a star-forming region \citep{H2D+_SOFIA_2014,HD2+_SOFIA_2017}. 

High degrees of methanol deuteration have been found in several hot corinos (which are the warm and dense inner parts of low-mass star-forming regions), including IRAS 16293$-$2422~B \citep{deuteration_16293_2018, deuterated_CH3OH_2019}. Furthermore, enhanced methanol deuteration has been demonstrated in the cold envelope of the low-mass Class 0 source L483 \citep{L483_3mm-survey_2019} and several starless prestellar cores, including L1544 \citep{L1544_deuteration_2019, Ambrose_2021, Lin_2023}. However, the deuterium enrichment in methanol is less pronounced in high-mass star-forming regions \citep{deuteration_NGC6334_2018,methanol-deuteration_no-CH3OD_2022} and even less so if the high-mass star-forming regions reside in the Galactic center, such as Sagittarius~B2(N2) \citep{EMoCA_with-D_2016}.

The rotational spectra of CH$_3$OD and other methanol isotopologs were first observed in the laboratory 
in the 1950s, and the initial goal was to determine their molecular structure \citep{MeOH-isos_1-0_I_1955}. 
Similar measurements were carried out by \citet{MeOH-isos_1-0_II_1956}, who also evaluated the height of the 
potential barrier to internal rotation based on CH$_3$OD data at $371 \pm 5$~cm$^{-1}$. 
\citet{CH_D3OH_D_rot_1968} published the first extensive study of its rotational spectrum in the millimeter wave region by investigating the torsion-rotation interaction in the methanol isotopologs CH$_3$OH, CD$_3$OH, and CH$_3$OD up to 200~GHz. The compilation of \citet{MeOH-RRF_1986} contained some unpublished CH$_3$OD data near 90~GHz taken by Lovas and Suenram in 1978. Additional measurements in the 14$-$92~GHz range were subsequently published by \citet{CH3OD_rot_dip_1980}, who also determined the dipole moment components through Stark effect measurements. 
The dipole moment components were redetermined shortly thereafter \citep{CH3OD_dip_1982}. 
\citet{CH3OD_rot_1988} expanded assignments of CH$_3$OD in the ground torsional state $\varv_{\rm t} = 0$ 
well into the submillimeter region. Some time later, \citet{CH3OD_rot_1993} made further assignments, 
including several in $\varv_{\rm t} = 1$. Two subsequent studies by \citet{CH3OD_rot_2000} and \citet{CH3OD_rot_2003} 
extended assignments to $\varv_{\rm t} = 2$. Both studies benefited from far-infrared laboratory measurements 
\citep[e.g.,][]{CH3OD_torsion_1997,CH3OD_FIR_vt0_1_1998,CH3OD_FIR_vt2_1999,CH3OD_FIR_vt2_2000}. 
In addition, several high-resolution infrared studies have been published. Important for our investigations are the works on the CO stretching band at 1042.7~cm$^{-1}$ \citep{CH3OD_CO-stretch_1994} and on the COD bending mode 
at 863.2~cm$^{-1}$ with a less detailed account of its hot band \citep{CH3OD_OD-bend_2013}. 
These two publications indicate interactions between the CO stretching state, the combination state 
of the COD bending with one quantum of the torsion, and $\varv_{\rm t} = 4$. 
In the course of our investigation, a report appeared on millimeter to far-infrared spectra of CH$_3$OD 
with a redetermination of its dipole moment components \citep{CH3OD_rot_FIR_dip_2021}. 

We have embarked on a program to extensively study various methanol isotopologs in low-lying torsional states in order 
to develop line lists with reliable positions and line strengths for astronomical observations and to 
investigate the intricate vibration-torsion-rotation interactions in their spectra. 
After our first reports on CD$_3$OH \citep{CD3OH_rot_2022} and CD$_3$OD \citep{CD3OD_rot_2023}, 
we have turned our attention to CH$_3$OD. We performed new measurements in the millimeter and submillimeter ranges 
to expand the frequency range with microwave accuracy up to 1.35~THz. 
The new data were combined in particular with previously published far-infrared measurements 
to form the final dataset involving rotational quantum numbers up to $J = 51$ and $K = 18$. 
A fit within the experimental errors was obtained for the ground and first excited torsional states of 
CH$_3$OD by employing the so-called rho-axis-method. 

We generated a line list that is based on our present results, which we applied to a reanalysis of CH$_3$OD in ALMA data of the Protostellar Interferometric Line Survey \citep[PILS,][]{PILS_2016} of the deeply embedded protostellar system IRAS 16293$-$2422. The new spectroscopic information leads to a lower column density in this source in comparison to the earlier determinations. This has major implications for our understanding of deuteration in the two functional groups of methanol.

The rest of the manuscript is organized as follows. Section~\ref{exptl} provides details on our 
laboratory measurements. The theoretical model, spectroscopic analysis, and fitting results are presented 
in Sections~\ref{spec_backgr} and \ref{lab-results}. Section~\ref{CH3OD_I16293B} describes our astronomical 
observations and the results of our present CH$_3$OD analysis, while Section~\ref{conclusion} provides 
the conclusions of our current investigation.

\section{Experimental details}
\label{exptl}
\subsection{Rotational spectra at the Universit{\"at} zu K{\"o}ln}

The spectral recordings at the Universit{\"a}t zu K{\"o}ln were carried out at room temperature 
using two different spectrometers. Pyrex glass cells of different lengths and with 
an inner diameter of 100~mm were employed. The cells were equipped with Teflon windows 
below $\sim$500~GHz; high-density polyethylene was used at higher frequencies. 
A commercial sample of CH$_3$OD (Sigma-Aldrich) was employed at initial pressures 
of 1.5 to 2.0~Pa. Minute leaks in the cells required a refill after several hours because 
of the slowly increasing pressure. These leaks caused some D-to-H exchange despite conditioning 
of the cells with higher pressures of CH$_3$OD prior to the measurements.
The resulting lines of CH$_3$OH did not pose any problem in the analyses because 
they can be easily identified from the work of \citet{CH3OH_rot_2008}. 

Both spectrometer systems used Virginia Diode, Inc. (VDI), frequency multipliers 
driven by Rohde \& Schwarz SMF~100A microwave synthesizers as sources. Schottky diode 
detectors were utilized below $\sim$500~GHz, whereas liquid He-cooled InSb 
bolometers (QMC Instruments Ltd) were applied between $\sim$500 and 1346~GHz.
Frequency modulation was used throughout, and the demodulation at $2f$ caused an isolated 
line to appear close to a second derivative of a Gaussian.

A double pass cell of 5~m in length was used to cover the 155$-$510~GHz range. 
Further information on this spectrometer is available elsewhere \citep{OSSO_rot_2015}. 
We achieved frequency accuracies of 5~kHz for the best lines with this spectrometer in a study 
of 2-cyanobutane \citep{2-CAB_rot_2017}, which exhibits a much richer rotational spectrum. 
We employed a setup with a 5-m single pass cell to cover 494 to 750~GHz, 760 to 
1093~GHz, and several sections of the 1117 to 1346~GHz region. Additional information on 
this spectrometer system is available in \citet{CH3SH_rot_2012}. We were able to achieve 
uncertainties of 10~kHz and even better for very symmetric lines with very good 
signal-to-noise (S/N) ratios, as demonstrated in recent studies on excited vibrational 
lines of CH$_3$CN \citep{MeCN_up2v4eq1_etc_2021} and on isotopic 
oxirane \citep{c-C2H4O_rot_2022,c-C2H3DO_rot_2023}. 
Uncertainties of 10, 20, 30, 50, 100, and 200~kHz were assigned in the present study, 
depending on the symmetry of the line shape, the S/N, and the frequency range. 
The smallest uncertainties above 1.1~THz were 50~kHz. 

\subsection{Rotational spectra at IRA NASU}

The measurements of the CH$_3$OD spectrum at the Institute of Radio Astronomy (IRA) of the National Academy of Sciences of Ukraine (NASU) were performed in the frequency ranges 34.4$-$183~GHz and 234$-$420~GHz using an automated synthesizer-based millimeter wave spectrometer \citep{Alekseev2023}. This instrument belongs to a class of 
absorption spectrometers and uses a set of backward wave oscillators (BWO) to cover 
the frequency range from 34.4 to 183~GHz, allowing further extension to the 234$-$420 GHz range 
employing a solid state tripler from VDI. The frequency of the BWO probing signal was stabilized 
by a two-step frequency multiplication of a reference synthesizer in two phase-lock-loop stages. 
A commercial sample of CH$_3$OD was used, and all measurements were carried out at room temperature, 
with sample pressures providing line widths close to the Doppler-limited resolution (about 2~Pa). 
The recorded spectrum contains numerous CH$_3$OH lines because of the relatively fast D-to-H exchange, 
as was observed at the Universit{\"a}t zu K{\"o}ln. Estimated uncertainties for measured line frequencies 
were 10, 30, and 100~kHz, depending on the observed S/N. 

\section{Spectroscopic properties of CH$_3$OD and our theoretical approach}
\label{spec_backgr}

The theoretical approach that we employed in the present study is the so-called rho-axis-method (RAM), 
which has proven to be the most effective approach so far in treating torsional large-amplitude motions 
in methanol-like molecules. The method is based on the work of \citet{Kirtman:1962}, 
\citet{CH_D3OH_D_rot_1968}, and \citet{Herbst:1984} and takes its name from the choice 
of its axis system \citep{Hougen:1994}. In RAM, the $z$ axis is coincident with the $\rho$ vector, 
which expresses the coupling between the angular momentum of the internal rotation $p_{\alpha}$ and 
that of the global rotation $J$. We employed the RAM36 code \citep{Ilyushin:2010,Ilyushin:2013}, 
which was successfully used in the past for a number of near-prolate tops with rather high $\rho$ and $J$ values 
(see, e.g., \citet{Smirnov:2014}, \citet{Motiyenko:2020}, \citet{Zakharenko:2019}, and \citet{Bermudez:2022}) and in particular for the CD$_3$OH and CD$_3$OD isotopologs of methanol \citep{CD3OH_rot_2022, CD3OD_rot_2023}. 
The RAM36 code uses the two-step diagonalization procedure of \citet{Herbst:1984}, 
and in the current study, we kept 41 torsional basis functions at the first diagonalization step 
and 11 torsional basis functions at the second diagonalization step. 

The OD-deuterated methanol, CH$_3$OD, is a nearly prolate top ($\kappa \approx -0.966$) with 
a rather high coupling between internal and overall rotations in the molecule ($\rho \approx 0.699$). Its 
torsional potential barrier $V_3$ is about 366~cm$^{-1}$. The torsional problem in CH$_3$OD corresponds 
to an intermediate barrier case \citep{RevModPhys.31.841} with the reduced barrier $s = 4V_3/9F$ $\sim$9.3, 
where $F$ is the rotation constant of the internal rotor. In comparison to the parent isotopolog, 
CH$_3$OD has somewhat smaller rotational parameters: $A \approx 3.68$~cm$^{-1}$, $B \approx 0.783$~cm$^{-1}$, and 
$C \approx 0.733$~cm$^{-1}$ in CH$_3$OD versus $A \approx 4.25$~cm$^{-1}$, $B \approx 0.823$~cm$^{-1}$, and
$C \approx 0.792$~cm$^{-1}$ in CH$_3$OH \citep{CH3OH_rot_2008}.  The angle between the RAM $a$-axis 
and the principal-axis-method (PAM) $a$-axis is 0.55$^{\circ}$, which is significantly larger than 
the corresponding angle of 0.07$^{\circ}$ in the parent methanol isotopolog. 
This larger angle in combination with its higher asymmetry ($\kappa \approx -0.966$ in CH$_3$OD versus 
$\kappa \approx -0.982$ in CH$_3$OH) leads to a situation where the labeling scheme after the second 
diagonalization step based on searching for a dominant eigenvector component starts to fail for 
some eigenvectors at $J \approx 24$. That is why we employed a so-called combined labeling scheme, 
where we used a dominant eigenvector component ($\geq 0.8$), if it exists, and we searched for similarities 
in the basis-set composition between the current eigenvector and the torsion–rotation eigenvectors 
belonging to the previous $J$ value and assigned the level according to the highest similarity found if 
a dominant eigenvector component is absent. This approach has already been applied successfully in the case 
of the CD$_3$OD study \citep{CD3OD_rot_2023}, where a more detailed description may be found. 
Further details of this labeling approach for torsion–rotation energy levels in low-barrier molecules based 
on similarities in basis-set composition of torsion–rotation eigenvectors of adjacent $J$ can be found 
in \citet{ILYUSHIN2004}.  

The energy levels in our fits and predictions are labeled by the free rotor quantum number $m$, 
the overall rotational angular momentum quantum number $J$, and a signed value of $K_a$, which is the axial 
$a$-component of the overall rotational angular momentum $J$. In the case of the A symmetry species, 
the $+/-$ sign corresponds to the so-called parity designation, which is related to the A1/A2 symmetry species 
in the group $G_6$ \citep{Hougen:1994}. The signed value of $K_a$ for the E symmetry species reflects the fact 
that the Coriolis-type interaction between the internal rotation and the global rotation causes levels with $|K_a| > 0$ 
to split into a $K_a > 0$ level and a $K_a < 0$ level. We also provide $K_c$ values for convenience, but they are 
simply recalculated from the $J$ and $K_{a}$ values: $K_{c} = J - |K_{a}|$ for $K_{a} \geq 0$ 
and $K_{c} = J - |K_{a}| + 1$ for $K_{a} < 0$. The $m$ values 0, $-$3, 3 / 1, $-$2, and 4 
correspond to A/E transitions of the $\varv_{\rm t} = 0$, 1, and 2 torsional states, respectively. 

\section{Spectroscopic results}
\label{lab-results}

We started our analysis from the microwave part of the dataset available in Tables~2 and 3 of 
\citet{CH3OD_rot_2003}, which consists of 994 $\varv_{\rm t} \leq 2$ microwave transitions 
ranging up to $J_{\rm max} = 21$ and $K_{\rm max} = 9$ augmented by the far-infrared measurements 
available in Table 2 of \citet{CH3OD_FIR_vt2_2000}. As a first step, we analyzed this combined 
dataset using the RAM36 program \citep{Ilyushin:2010,Ilyushin:2013} and used the resulting fit 
as the starting point for our assignments. New data were assigned starting with the Kharkiv measurements, 
which were done in parallel for the three lowest torsional states of CH$_3$OD  $\varv_{\rm t} = 0$, 1, and 2. 
Submillimeter wave and terahertz measurements from K{\"o}ln were assigned subsequently based on our new results. 
The assignment process was performed in a usual bootstrap manner, with numerous cycles of refinement 
of the parameter set while gradually adding the new data. Whenever it was possible, we replaced 
the old measurements from \citet{CH3OD_rot_2003} and references therein with the more accurate new ones. In the best case, this gave us an improvement in measurement uncertainty from 100 kHz to 10 kHz, whereas in the worst case, a reduction of uncertainty from 50 kHz to 30 kHz was achieved. At the same time, as we already did in the cases of the CD$_3$OH \citep{CD3OH_rot_2022} and CD$_3$OD 
\citep{CD3OD_rot_2023} studies, we decided to keep the two measured values for the same transition in the fits
from the Kharkiv and K{\"o}ln spectral recordings in that part of the frequency range where the measurements 
from the two laboratories overlap (154$-$183~GHz and 234$-$420~GHz). A rather good agreement within 
the experimental uncertainties was observed for this limited set of duplicate new measurements. 
Finally, at an advanced stage of our analysis, the far-infrared data from \citet{CH3OD_rot_FIR_dip_2021} 
were added to the fit. 

In the process of searching for the optimal set, it became evident that the $\varv_{\rm t}=2$ 
torsional state poses some problems with fitting. The strong influence of intervibrational interactions 
arising from low-lying small-amplitude vibrations in CH$_3$OD (see, for example, 
\citet{CH3OD_CO-stretch_1994, CH3OD_OD-bend_2013}), which then propagate down through numerous 
intertorsional interactions, is a possible explanation for these problems. We encountered similar problems 
with CD$_3$OH \citep{CD3OH_rot_2022} and CD$_3$OD \citep{CD3OD_rot_2023}. In the future, we plan to account explicitly for the above-mentioned intervibrational interactions, and with this aim in mind, new measurements of the CH$_3$OD IR spectrum between 500 and 1200~cm$^{-1}$ were carried out at the Technische Universität Braunschweig. These measurements were not used in the present investigation. Therefore, the details of these new measurements will be presented in due course when the new data will be included in our analysis of intervibrational interactions. In the meantime, the difficulties in fitting the $\varv_{\rm t}=2$ data within experimental uncertainties prompted us to limit our analysis mainly to the ground and first excited torsional states, thus providing reliable rest frequencies for radio astronomical observations of CH$_3$OD. At the final stage of the model refinement, our fit included, besides the ground and first excited torsional states 
of CH$_3$OD, the lowest three $K$ series for the A and E species in $\varv_{\rm t}=2$ in order to obtain a better constraint 
of the torsional parameters in the Hamiltonian model. These $\varv_{\rm t}=2$ $K$ levels should be the least affected 
by the intervibrational interactions arising from low-lying small-amplitude vibrations. In the case of CH$_3$OD, 
this corresponds to $K = -1,-2, 3$ for the E species in $\varv_{\rm t}=2$ and to $K = -1, 0, 1$ for the A species. 

Our final CH$_3$OD dataset contains 4758 far-infrared and 10163 microwave line frequencies. Due to blending, these 14921 
measured frequencies correspond to 16583 transitions with $J_{\rm max} = 51$ and $K_a \leqslant 18$. 
Taking into account duplicate measurements mentioned above, our final dataset represents 15049 unique transitions in the fit. A Hamiltonian model consisting of 134 parameters provided a fit with a weighted root mean square (wrms) deviation of 0.85, which was selected as our "best fit" for this paper. The 134 molecular parameters from our final fit are given in Table~\ref{tbl:ParametersTable} (Appendix A). The numbers of the terms in the model distributed between the orders $n_{\rm op}$ = 2, 4, 6, 8, 10, 12 are 7, 22, 46, 39, 16, 4, respectively, which is consistent with the limits of determinable parameters of 7, 22, 50, 95, 161, and 252 for these orders, as calculated from the differences between the total number of symmetry-allowed Hamiltonian terms of order $n_{\rm op}$ and the number of symmetry-allowed contact transformation terms of order $n_{\rm op} - 1$ when applying the ordering scheme of \citet{Nakagawa:1987}. The final set of the parameters converged perfectly in all three senses: (i) the relative change in the wrms deviation of the fit at the last iteration was about $\sim 5 \times 10^{-7}$; (ii) the corrections to the parameter values generated at the last iteration were less than $\sim$10$^{-3}$ of the calculated parameter confidence intervals; and (iii) the changes generated at the last iteration in the calculated frequencies were less than 1~kHz, even for the far-infrared data. 


\begin{figure}
\centering
\includegraphics[width=9cm,angle=0]{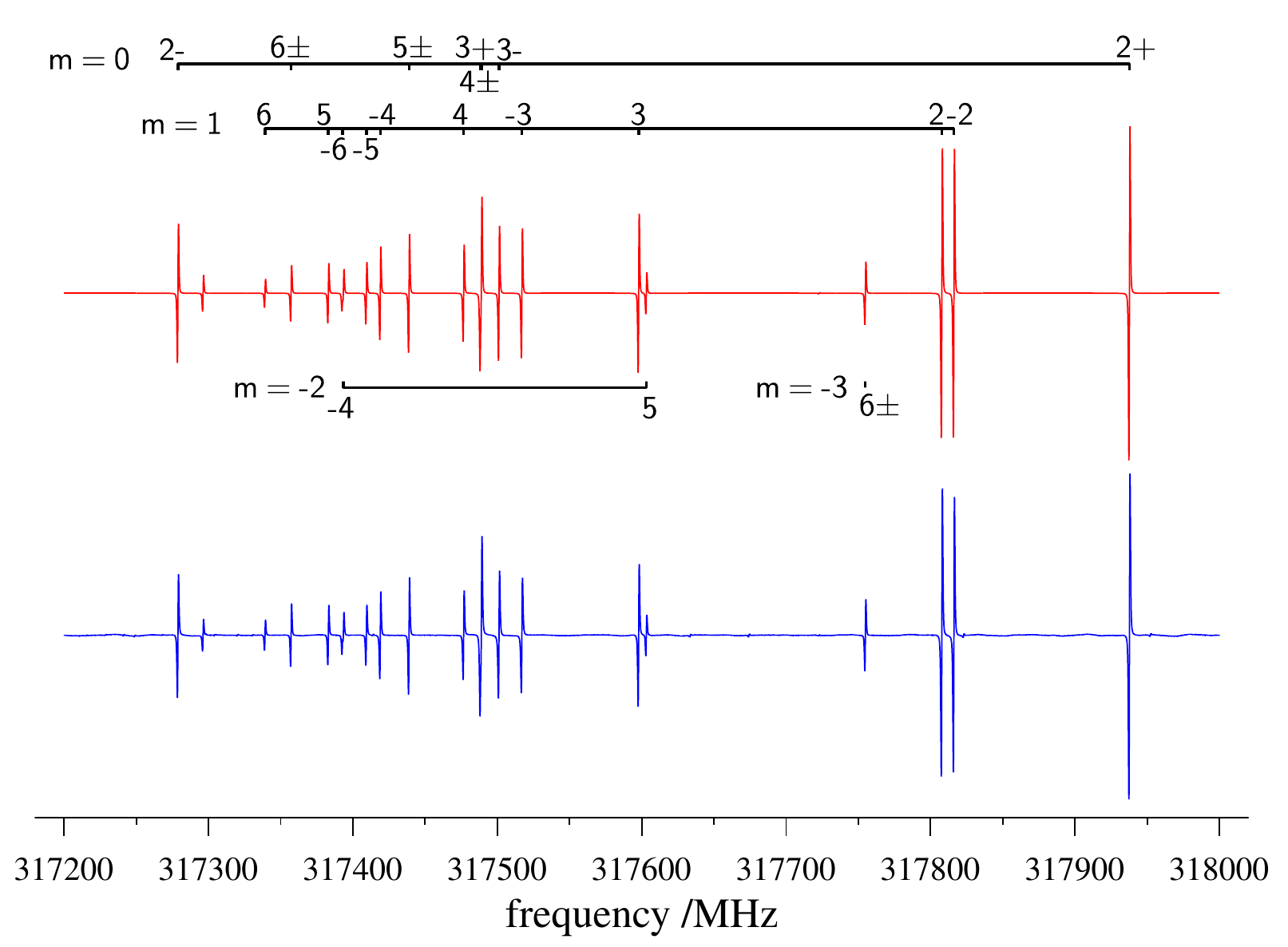}

\caption{Portion of the CH$_3$OD microwave spectrum dominated by the $J = 7 \leftarrow 6$ $R$-branch of transitions 
in the 317.2$-$318.0~GHz range.  The observed spectrum is shown in the lower panel, and the calculated one is in the upper panel. The spectrum was recorded in the first derivative detection mode of the Kharkiv spectrometer. The $K$ quantum numbers for the $m=0,1$ ($\varv_{\rm t} = 0$) transitions are given above, and for the $m=-2,-3$ ($\varv_{\rm t} = 1$) transitions, they are given below the simulated spectrum in the upper panel. We observed that the experimental frequencies and the  intensity pattern are rather well reproduced by our model for the spectral features dominating this frequency range.}
\label{fig_MMW_Kharkiv}
\end{figure}



\begin{figure}
\centering
\includegraphics[width=9cm,angle=0]{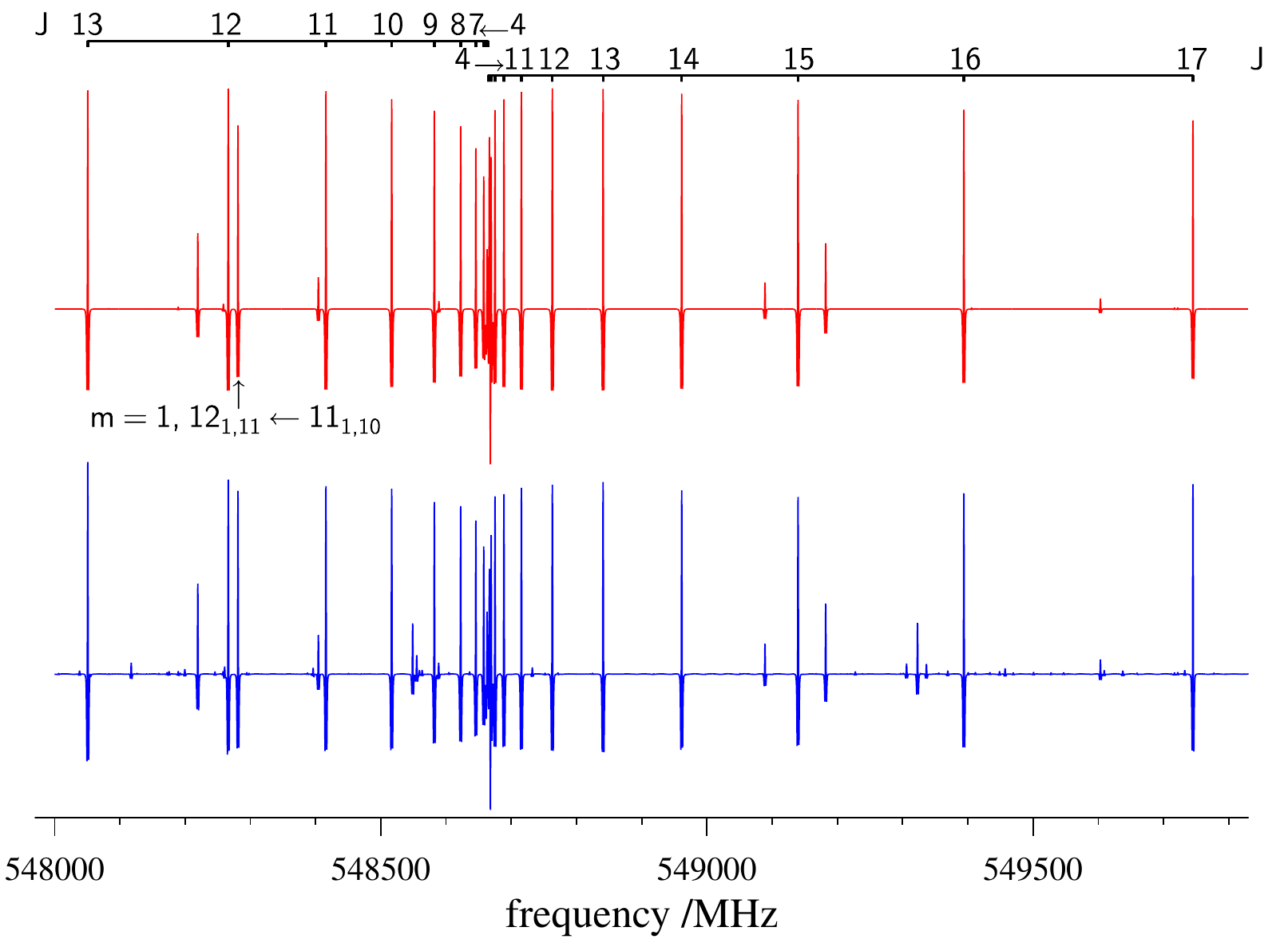}

\caption{Portion of the CH$_3$OD microwave spectrum dominated by the $m = 0$ $K= 4\pm \leftarrow 3\mp $ $Q$-branch transitions in the 548.0$-$549.8~GHz range. The observed spectrum is shown in the lower panel, and the calculated one is in the upper panel. The spectrum was recorded in the second derivative detection mode of the K{\"o}ln spectrometer. The $J$ quantum numbers of the $Q$-branch transitions given at the top decrease with frequency for the $K= 4+ \leftarrow 3-$ transitions and increase with frequency for the $K= 4- \leftarrow 3+$ transitions. The arrows in progressions of $J$ quantum numbers shown in the top panel indicate the successive $J$ values (from four to seven for $K= 4+ \leftarrow 3-$ and from four to 11 for $K= 4- \leftarrow 3+$) that we were not able to show separately within the selected scale of the figure.  Also, we have marked the $m = 1$, $12_{1,11} \leftarrow 11_{1,10}$ $R$-type transition, which has a comparable intensity with dominating Q-branch transitions in this range. We observed that the experimental frequencies and the  intensity pattern are rather well reproduced by our model for the spectral features dominating this frequency range.}
\label{fig_SMMW_Koln}
\end{figure}


A summary of the quality of this fit is given in Table~\ref{tbl:statisticInf}. In the left part of 
Table~\ref{tbl:statisticInf}, the data are grouped by measurement uncertainty, and all data groups are fit within 
experimental uncertainties. We observed the same good agreement in the right part of Table~\ref{tbl:statisticInf}, 
where the data are grouped by torsional state. The overall wrms deviation of the fit is 0.85. 
A further illustration of the rather good agreement between the observed and the calculated line positions and intensities from our final Hamiltonian model in the spectrum of CH$_3$OD can be seen in Figs.~\ref{fig_MMW_Kharkiv} and \ref{fig_SMMW_Koln}. 


\begin{table*}[]
\centering
\caption{\label{tbl:statisticInf} Overview of the dataset and the fit quality.}
\begin{tabular}{lrl|lrl}
\hline
\multicolumn{3}{c|}{By measurement uncertainty}& \multicolumn{3}{c}{By torsional state} \\ 
\cline{1-6}
 
\multicolumn{1}{c}{Unc.$^a$} & \multicolumn{1}{c}{$\#^b$} & \multicolumn{1}{c|}{rms$^c$} 
& \multicolumn{1}{c}{$\varv_{\rm t}^d$} & \multicolumn{1}{c}{$\#^b$} & \multicolumn{1}{c}{wrms$^e$} \\

\cline{1-6}

0.010~MHz & 5198 & 0.0085~MHz & $\varv_{\rm t}=0 \leftarrow 0$ & 6847 & 0.82 \\
0.020~MHz &  932 & 0.0169~MHz & $\varv_{\rm t}=1 \leftarrow 1$ & 5694 & 0.88 \\
0.030~MHz & 1095 & 0.0246~MHz & $\varv_{\rm t}=2 \leftarrow 2$ &  363 & 0.89 \\
0.050~MHz & 1224 & 0.0389~MHz & $\varv_{\rm t}=1 \leftarrow 0$ & 3330 & 0.94 \\
0.100~MHz & 1086 & 0.0765~MHz & $\varv_{\rm t}=2 \leftarrow 1$ &  346 & 0.87 \\
0.200~MHz &  628 & 0.1382~MHz & $\varv_{\rm t}=1 \leftarrow 2$ &    3 & 0.62 \\
$2\times 10^{-4}$~cm$^{-1}$ & 4758 & $1.8\times 10^{-4}$~cm$^{-1}$ &   &  & \\ \hline

\end{tabular}
\tablefoot{$^{a}$ Estimated measurement uncertainties for each data group. $^{b}$ Number of lines (left part) or transitions (right part) of each category in the least-squares fit. We note that due to blending, 14921  measured line frequencies correspond to 16583  transitions in the fit, which in turn due to the presence of duplicate measurements represent 15049 unique transitions in the fit.  $^{c}$ Root mean square deviation of corresponding data group. $^{d}$ Upper- and lower-state torsional quantum number $\varv_{\rm t}$. $^{e}$ Weighted root mean square deviation of the corresponding data group.}
\end{table*}


We calculated a CH$_3$OD line list in the ground and first excited torsional states from the parameters of our final 
Hamiltonian model for radio astronomical observations. The dipole moment function of \citet{MEKHTIEV1999171} 
was employed in our calculations where the values for the permanent dipole moment components of CH$_3$OH were replaced 
by appropriate ones for CH$_3$OD; $\mu_a = 0.8343$~D and $\mu_b = 1.4392$~D were taken from \citet{CH3OD_rot_FIR_dip_2021}. 
The permanent dipole moment components were rotated from the principal axis system to the rho axis system 
of our Hamiltonian model. As in the cases of CD$_3$OH \citep{CD3OH_rot_2022} and CD$_3$OD \citep{CD3OD_rot_2023}, 
the list of CH$_3$OD transitions includes information on transition quantum numbers, transition frequencies, 
calculated uncertainties, lower-state energies, and transition strengths. As already mentioned, 
we labeled torsion-rotation levels by the free rotor quantum number $m$, the overall rotational angular momentum 
quantum number $J$, a signed value of $K_a$, and $K_c$. To avoid unreliable extrapolations far beyond the quantum number 
coverage of the available experimental dataset, we limited our predictions by $\varv_{\rm t} \leq 1$, $J \leq 55$ and 
$|K_{a}| \leq 21$. 
The calculations were done up to 2.0~THz. Additionally, we limited our calculations to transitions for which calculated 
uncertainties are less than 0.1~MHz. The lower-state energies are given referenced to the $J = 0$ A-type $\varv_{\rm t} = 0$ 
level. In addition, we provide the torsion-rotation part of the partition function $Q_{\rm rt}$(T) of CH$_3$OD calculated 
from first principles, that is, via direct summation over the torsion-rotational states. The maximum $J$ value is 65 
for this calculation, and $n_{\varv_{\rm t}} = 11$ torsional states were taken into account. The calculations, as well as 
the experimental line list from the present work, can be found in the online supplementary material of this article 
and will also be available in the Cologne Database for Molecular Spectroscopy \citep[CDMS,][]{CDMS_2001, CDMS_2005, CDMS_2016}.

\section{CH$_{3}$OD in IRAS~16293-2422~}
\label{CH3OD_I16293B}

\begin{figure}
	\centering
	\includegraphics[width=\hsize]{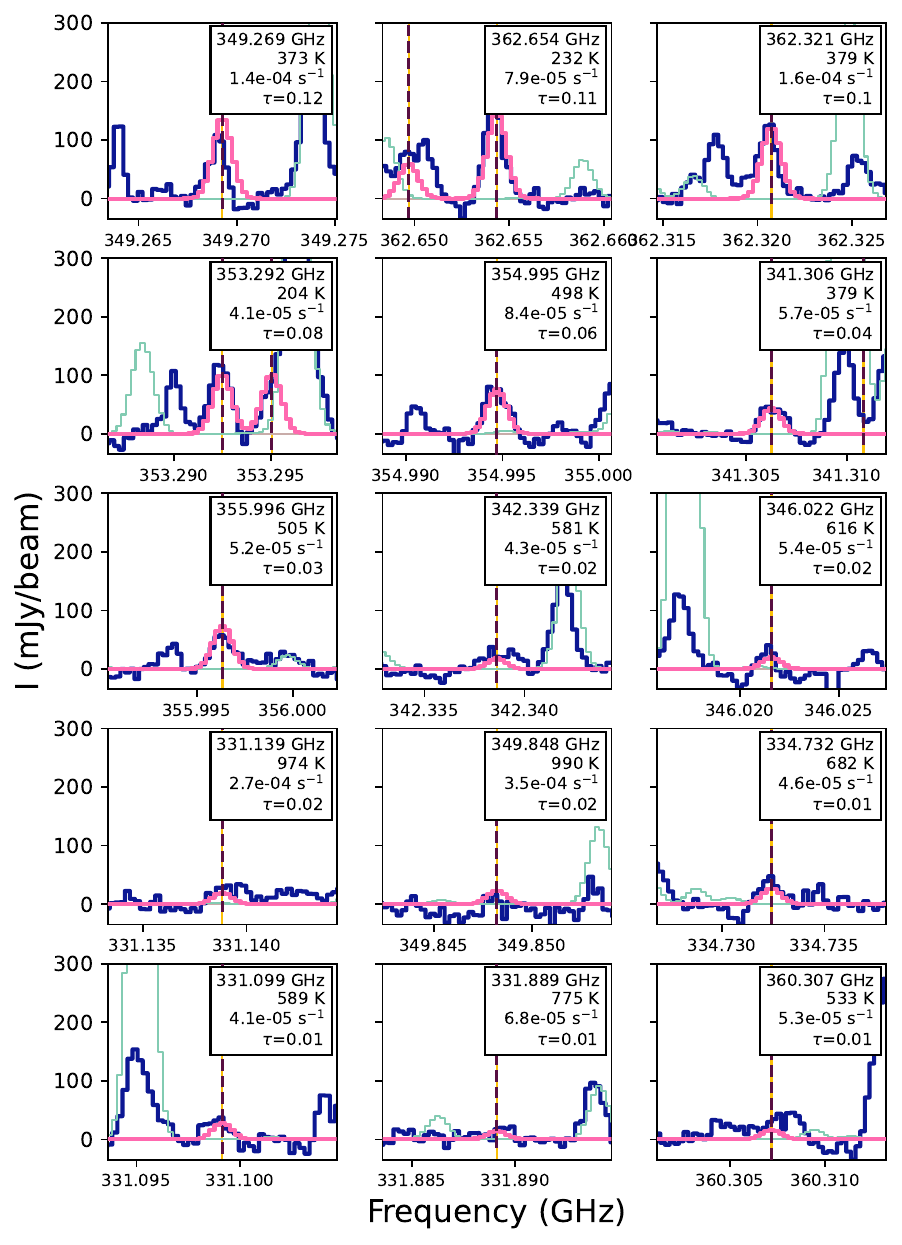}
	\caption{Selection of $15$ detected lines of CH$_{3}$OD ordered by their line opacity ($\tau$) from top left to bottom right. The observed spectrum is in dark blue, the reference spectrum (corresponding to all other
species identified thus far in PILS) is in turquoise, and the best-fitting synthetic spectrum of CH$_{3}$OD is in pink. The rest frequency, $E_{\text{up}}$ (K), $A_{ij}$ (s$^{-1}$), and $\tau$ (for the best-fitting parameters) are shown in the right corner of each panel. The rest frequency is indicated with a vertical dashed line, and the filled yellow region corresponds to the uncertainty on that line frequency. The lines at $355.996$, $334.732$, and $331.099$~GHz are in fact unresolved asymmetry doublets. Only the panels corresponding to $341.306$, $355.996$, $342.339$, $346.022$, $334.732$, $331.099$, $331.889$, and $360.307$~GHz illustrate lines that have been included in constraining the best-fit synthetic spectrum, as the others become (partially) optically thick ($\tau>0.1$) at $T_{\text{ex}}=50$~K, $300$~K, or both.}
	\label{fig:CH3OD_panel_plot}
\end{figure}

\begin{figure*}
	\centering
	\includegraphics[width=\hsize]{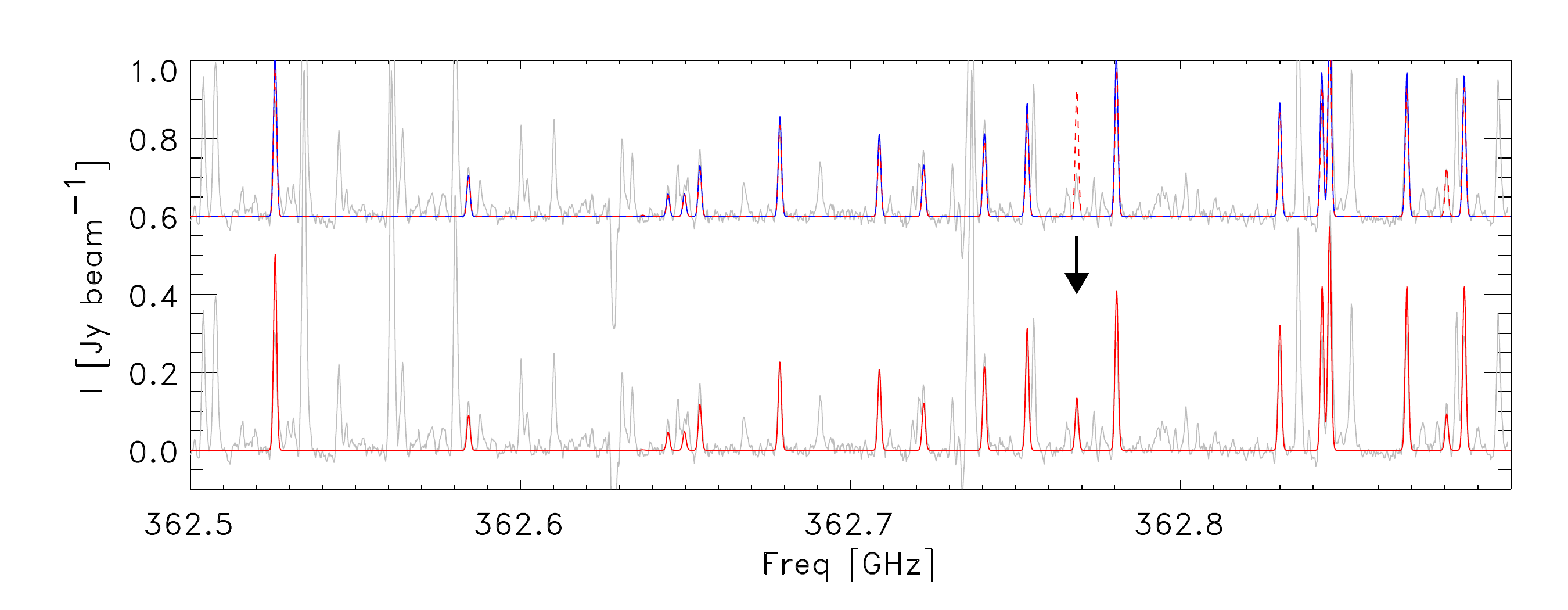}

	\caption{Comparison between the observed PILS spectrum and the synthetic spectra obtained using the (old) spectroscopic data utilized in \citet{deuteration_16293_2018} (in blue) and the current spectroscopy (in red). The two synthetic spectra at an offset of $0.6$~Jy~beam$^{-1}$ correspond to the prediction from the fit of the 2018 paper (solid blue curve) and current spectroscopy (dashed red curve), adopting the excitation temperature of $300$~K from \citet{deuteration_16293_2018} and the column density from that same paper corrected for the factor of four error discussed in Section~\ref{CH3OD_I16293B} ($N=4.5\times10^{16}$~cm$^{-2}$). These two synthetic spectra are virtually identical except for two high excitation transitions at $362.7685$ and $362.8805$~GHz, where the new spectroscopic model demonstrates one clearly overpredicted transition (marked by an arrow). The synthetic spectrum at an offset of $0$~Jy/beam  correspond to the prediction with the lower inferred excitation temperature from the analysis summarized in Table~\ref{tbl:N_Tex}, shown with a red curve. As can be seen, this spectrum matches well the high excitation $362.7685$ and $362.8805$~GHz transitions, which were not included in the old spectroscopy. These two lines come out at appropriate intensities in comparison to the observations within the uncertainties of the model.}
	
	\label{fitexample}
\end{figure*}

The new spectroscopic calculations were used to reanalyze the emission lines of CH$_{3}$OD in data from the Protostellar Interferometric Line Survey (PILS;\footnote{\url{http://youngstars.nbi.dk/PILS/}} project-id: 2013.1.00278.S, PI: Jes K. J\o{}rgensen). PILS represents an unbiased spectral line survey of the Class~0 protostellar system IRAS~16293-2422 using ALMA and covering the frequency range from $329$ to $363$~GHz. The observations target the region of IRAS~16293-2422, including its two primary components "A" and "B" that show abundant lines of complex organic molecules at an angular resolution of  $\sim0.5''$ and a spectral resolution of $\sim0.2$~km~s$^{-1}$. Toward a position slightly offset from the "B" component of the system (RA, Dec (J2000) of $16^{\text{h}}32^{\text{m}}22.58^{\text{s}}$, $-24^{\circ}28\arcmin32.80\arcsec$), the lines are intrinsically narrow, making it an ideal hunting ground for new species. Several complex organic molecules and their isotopologs have already been identified there, including deuterated isotopologs of CH$_{3}$OH, namely CH$_{2}$DOH and CH$_{3}$OD \citep{deuteration_16293_2018}, CHD$_{2}$OH \citep{CHD2OH_catalog_2022}, and CD$_{3}$OH \citep{CD3OH_rot_2022}. A tentative detection of several CD$_{3}$OD lines has also been reported at this position \citep{CD3OD_rot_2023}. The full details on the PILS dataset and its reduction are available in \citet{PILS_2016}.

The reanalysis of CH$_{3}$OD was conducted by fitting synthetic spectra to the observations with calculations that assume that the excitation of the molecule is characterized by local thermodynamical equilibrium (LTE), which is reasonable at the densities on the spatial scales probed by PILS (namely, H$_{2}$ number density $>3 \times 10^{10}$ cm$^{-3}$, which was demonstrated to result in deviations between the excitation and kinetic temperatures of less than $15\%$; Section~$5.1$ of \citealt{PILS_2016}). For all CH$_{3}$OD lines, the velocity offset relative to the local standard of rest matches the canonical $2.7$~km~s$^{-1}$ at this position, and its line widths are well fit by the typical $1$~km~s$^{-1}$ full width half maximum (FWHM). Beam size and source size were both fixed to $0.5''$ and were both assumed to have Gaussian distributions (i.e., beam-filling factor of $0.5$). The fitting methodology is based on the MCMC Python package \textsc{emcee} \citep{ForemanMackey2013},\footnote{\url{https://emcee.readthedocs.io/en/stable/}} and its application is described in detail in Section~$2.3$ of \citet{CHD2OH_catalog_2022}. There are in total $485$ lines of CH$_{3}$OD ($243$ in $v_{t}=0$ and $242$ in $v_{t}=1$) covered in the observed PILS frequency range, of which $480$ ($241$ in $v_{t}=0$ and $239$ in $v_{t}=1$) have unique rest frequencies. All covered (detected and non-detected) lines of CH$_{3}$OD were investigated for potential blending with already identified molecules in this source. All lines that have any level of blending were removed from further synthetic spectrum fitting. Optically thick lines of CH$_{3}$OD were also removed from further synthetic spectrum fitting. These are assumed to be lines that have $\tau>0.1$ either at $T_{\text{ex}}=50$ or $300$~K at $N=4.5\times10^{16}$~cm$^{-2}$ (the best-fit column density of CH$_{3}$OD derived in \citealt{deuteration_16293_2018}, corrected for the factor of four error discussed below). After the removal of the blended and optically thick lines, $97$ unique line frequencies remained, and they were then used for synthetic spectral fitting (this includes detected and non-detected lines). Out of the $97$ lines, $28$ were predicted to have a peak intensity greater than $14$~mJy~beam$^{-1}$ (and an integrated intensity greater than $3\sigma$ for $\sigma=4.5$~mJy~beam$^{-1}$~km~s$^{-1}$ and a line width of $1$~km~s$^{-1}$) for the subsequently derived best-fitting $T_{\text{ex}}$ and $N$ (Table~\ref{tbl:N_Tex}). In this way, non-blended, non-detected lines were also used to constrain the synthetic spectral fitting. The number of walkers and the parameter space setup used here match what was used for the analysis of CHD$_{2}$OH in \citet{CHD2OH_catalog_2022}. Only for the case of fitting CH$_{3}$OD in the $\varv_{\rm t}=0$ and $\varv_{\rm t}=1$ states together, the number of steps had to be increased from $1~000$ to $1~500$ to ensure proper convergence. For the second MCMC run of the computation, the mean acceptance fraction of the $300$ walkers is $70-71\%$ (independent of whether the $\varv_{\rm t}=0$ and $\varv_{\rm t}=1$ states were fitted together or separately), and the quality of the convergence is illustrated in the corner plot shown in Fig.~\ref{fig:CH3OD_corner} (Appendix B) for IRAS~16293-2422~B.



\begin{table}
\centering
\caption{Best-fitting column densities, excitation temperatures, and D/H ratios of CH$_{3}$OD toward IRAS~16293-2422~B.}
\label{tbl:N_Tex}
\centering
\begin{tabular}{l l}
\hline\hline
Best-fitting parameters                           & IRAS~16293-2422~B \\
\hline
Fitting $\varv_{\rm t} = 0$ and $1$ together              & \\
\hline
$T_{\text{ex}}$ (K)                               & $190\pm19$\\
$N($CH$_{3}$OD$_{\varv_{\rm t} = 0, 1})$ (cm$^{-2}$)      & $(3.25\pm0.65)\times10^{16}$\\
D/H of CH$_{3}$OH                                 & $(3.2\pm0.9)\times10^{-3}$\\
\# lines                                           &  $97$                       \\
$E_{\text{up}}$ range (K)                          &  $124-2969$                 \\
\hline
Fitting solely $\varv_{\rm t}=0$                        & \\
\hline
$T_{\text{ex}}$ (K)                               & $205\pm21$\\
$N($CH$_{3}$OD$_{\varv_{\rm t} = 0})$ (cm$^{-2}$)         & $(2.52\pm0.50)\times10^{16}$\\
D/H of CH$_{3}$OH                                 & $(2.5\pm0.7)\times10^{-3}$\\
\# lines                                           &  $46$                       \\
$E_{\text{up}}$ range (K)                          &  $124-2969$                 \\
\hline
Fitting solely $\varv_{\rm t}=1$                        & \\
\hline
$T_{\text{ex}}$ (K)                               & $163\pm16$\\
$N($CH$_{3}$OD$_{\varv_{\rm t} = 1})$ (cm$^{-2}$)         & $(4.01\pm0.80)\times10^{16}$\\
D/H of CH$_{3}$OH                                 & $(4.0\pm1.0)\times10^{-3}$\\
\# lines                                           &  $51$                       \\
$E_{\text{up}}$ range (K)                          &  $332-2287$                 \\
\hline
\end{tabular}
\tablefoot{Top part contains the best-fitting column density and excitation temperature of CH$_{3}$OD obtained during simultaneous fitting of these two parameters for both $\varv_{\rm t}=0$ and $1$ states together and the associated D/H ratio. We note that the statistical correction on the D/H ratio for the case of deuteration in the hydroxy group is one (Appendix~C of \citealt{CHD2OH_catalog_2022}; see also \citealt{Manigand2019}). The fitting has also been executed for the two $\varv_{\rm t}$ states separately, for which the corresponding derived column densities and D/H ratios are the middle and lower parts, respectively. The second column contains the values for the one-beam offset position in the SW direction from IRAS~16293-2422~B ($0.5\arcsec$ or $70$~au away from the dust peak of B). For the calculations of D/H ratios, the adopted methanol column densities is $(1.0\pm0.2)\times10^{19}$~cm$^{-2}$ \citep{deuteration_16293_2018}. The stated errors on $N$ and $T_{\text{ex}}$ are assumed $20\%$ and $10\%$ errors, respectively. The errors derived with the MCMC routine in this work are much smaller (we refer to \citet{CHD2OH_catalog_2022} for a discussion about that). For each of the three fitting scenarios, the number of lines with unique frequencies used for the fitting and the $E_{\text{up}}$ range spanned by these lines are also provided.}
\end{table}


Fitting the $\varv_{\rm t}=0$ and $1$ states simultaneously yields a best-fitting $T_{\text{ex}}=190\pm19$~K and $N=(3.25\pm0.65)\times10^{16}$~cm$^{-2}$ (Table~\ref{tbl:N_Tex}, Fig.~\ref{fig:CH3OD_panel_plot}). If the two $\varv_{\rm t}$ states are fitted separately, then it becomes apparent that the fit is driven to $T_{\text{ex}} < 200$~K based on the $\varv_{\rm t}=1$ lines. However, the $\varv_{\rm t}=0$ lines are best fit by $T_{\text{ex}} \approx200$~K, which, considering the error bars, cannot be firmly ascertained as stemming from a component of a different temperature. The differences in $N$ for the range of best-fitting $T_{\text{ex}}$ depending on whether the $\varv_{\rm t}$ states are fit together or separately are less than a factor of $1.6$. Hence, there is no reason to fit the $\varv_{\rm t}=0$ and $1$ states of CH$_{3}$OD separately in the PILS data. 

Using the new spectroscopic data significantly reduces the best-fitting $T_{\text{ex}}$ and $N$ for CH$_{3}$OD in comparison to the earlier estimates in \citet{deuteration_16293_2018} of $N=1.8\times10^{17}$~cm$^{-2}$ and $T_{\text{ex}}=300$~K, which corresponds to a reduction by a factor of $\sim$5.5 in $N$ and $\sim$1.6 in $T_{\text{ex}}$. Part of this difference in the column density is due to an incorrect coupling between the partition function for CH$_{3}$OD and the line intensities for the spectroscopic data utilized in \citet{deuteration_16293_2018}. For the partition function, that paper utilized a scaled version of the partition function for CH$_{3}^{18}$OH but did not take into account the factor for $g_{I}$. This factor was equal to one for the CH$_{3}$OD intensities from \citet{CH3OD_rot_1988}, but it was equal to four for the scaled partition function.\footnote{See explanation at \url{https://cdms.astro.uni-koeln.de/classic/predictions/description.html}; update November 2023.} Consequently, the modeled line intensities from the synthetic spectra were underestimated by a factor of four, and thus, the derived column density is overestimated by the same factor.

Another effect also comes into play through the derived excitation temperature: \citet{deuteration_16293_2018} derived a temperature of $321\pm33$~K from a rotation diagram fit (Fig.~$1$ in that paper) and categorized CH$_{3}$OD as being one of the species belonging to the "high-temperature" group ($300$ versus $125$~K) of species. However, the new spectroscopic data appear to drive the best-fit excitation temperature toward lower values. The reason for this is illustrated with an example in Fig.~\ref{fitexample} that shows a zoom-in on a specific frequency range between $362.5$ and $362.9$~GHz, which harbors several prominent CH$_{3}$OD transitions. 

The upper part of the figure shows a comparison of the best-fit synthetic spectrum to the PILS data using the spectroscopy utilized in \citet{deuteration_16293_2018} and the spectroscopy of this paper for an excitation temperature of $300$~K and a column density of $N=4.5\times10^{16}$~cm$^{-2}$ (the best-fit column density of CH$_{3}$OD derived in \citealt{deuteration_16293_2018}, corrected for the factor of four error mentioned above). As shown, the fits are very close except for one transition ($\varv_{\rm t}=0$, A-type, $22_{1-,22}-22_{0+,22}$) at $362.7685$~GHz, which has a high upper energy level that was not included in the old spectroscopy. This line with the high upper energy level of $563$~K is significantly overproduced, with a high excitation temperature of $300$~K, and consequently drives the fit toward a lower value. The spectral range also includes one $\varv_{\rm t}=1$ transition at $362.8805$~GHz ($E_{\text{up}}$ of $343$~K) that is included in the new spectroscopy, but not in the original spectroscopy utilized in \citet{deuteration_16293_2018}. The observed feature is well matched with the predictions from the $300$~K fit of the 2018 analysis (even though it was not included there). With the lower excitation temperature fit (lower panel of Fig.~\ref{fitexample}), its line strength is slightly below ($30\%$) the observed spectrum and the $300$~K fit, but it is still in agreement within the uncertainties. The lower synthetic spectrum corresponds to the excitation temperature lowered to the value of Table~\ref{tbl:N_Tex} and the best-fit column density from the current analysis.

On the other hand, an important caveat here is of course that it is likely that there are gradients in temperatures, column densities, and extents of the emission along the line of sight. The fact that the higher excitation transitions from the new data are overproduced may also reflect that they represent more compact emission with smaller filling factors compared to the beam. If higher excitation transitions stem from a region that is significantly smaller than the beam size ($0.5''$), then our synthetic spectrum fitting would overestimate the source size. This effect could be compensated by driving the fit to lower excitation temperatures. However, fitting lower and higher excitation transitions separately is not a solution because the lower excitation transitions would be excited in high-temperature regions as well. Such gradients in temperature and differences in the extents of the high versus low excited transitions would of course also affect the line optical thickness (underestimating it for the lines tracing more compact emission). However, considering the lines according to their optical thickness, it is mainly the lower excited transitions that are likely to become optically thick, and those, in fact, do show extended emission relative to the beam (see Fig.~2 in \citealt{deuteration_16293_2018} for the spatial distribution of various methanol isotopologs in lines with $E_{\text{up}}$ on the order of $150-300$~K). Without dedicated higher spatial resolution observations to use as constraints, it makes little sense to introduce more free parameters into the synthetic spectrum fitting. In any case, this discussion illustrates the importance of complete and updated spectroscopy, and it also emphasizes the need for using the comparisons between the observed spectra and synthetic models also including predictions for transitions that would not directly be predicted to be observed.

The newly determined best-fitting D/H ratio for CH$_{3}$OD is $(0.32\pm0.09)\%$. In contrast to what was previously thought, this implies that the D/H ratio in the hydroxy group of methanol does not match the D/H ratio in the methyl group of methanol. Including the statistical correction of three, the methyl group deuteration is $(2.4\pm0.72)\%$  \citep{deuteration_16293_2018}, while the newly determined hydroxy group deuteration is a factor of $\sim$7.5 lower. This results in the hydroxy group of methanol having the lowest D/H ratio of all molecules with a measured D/H ratio in IRAS~16293-2422~B thus far. This is consistent with laboratory experiments that have demonstrated that deuteration in the hydroxy group is a lot less efficient than deuteration in the methyl group \citep{Nagaoka2005, Nagaoka2007, Hidaka2009}; however, there are contrasts with the hydroxy group deuterated isotopologs of ethanol and formic acid with D/H ratios on the order of a few percent in IRAS~16293-2422~B \citep{deuteration_16293_2018}. This points to the fact that our understanding of the synthesis of complex organic molecules and their deuteration remains incomplete.

\section{Conclusions}
\label{conclusion}

We have carried out an extensive study of the torsion-rotation spectrum of CH$_3$OD 
using a torsion-rotation RAM Hamiltonian. The new microwave measurements were performed in the broad 
frequency range from 34.4~GHz to 1.35~THz. Transitions involving the $\varv_{\rm t}$ = 0, 1, and 2 
torsional states with $J$ up to 51 and $K_a$ up to 18 were assigned and analyzed in the current work. 
The second torsional state posed some problems in obtaining a fit within experimental uncertainties 
using our current model, as was the case in our earlier investigations of CD$_3$OH and CD$_3$OD. 
We suspect perturbations by intervibrational interactions, which arise from low-lying 
small-amplitude vibrations of CH$_3$OD and transfer down to lower torsional states via torsion-rotation interactions, as the main reason for this. Therefore, we concentrated our efforts on refining the theoretical model for the ground and 
the first excited torsional states only, as in our studies of CD$_3$OH and CD$_3$OD. 
We achieved a fit well within the experimental uncertainties, with a weighted rms deviation of 0.85 
for the dataset, which consists of 4758 far-infrared and 10163 microwave line frequencies. 

We carried out calculations of the ground and first excited torsional states' spectra on the basis of our results 
and used these calculations to reinvestigate CH$_3$OD in data from PILS, which is a spectral survey of the deeply embedded low-mass protostar IRAS 16293$-$2422 performed with ALMA. Both, $\varv_{\rm t}=0$ and $\varv_{\rm t}=1$ transitions are observed in these data. The new accurately determined value for the column density of CH$_3$OD is a factor of $\sim$5.5 lower than earlier estimates. This implies a D/H ratio of $(0.32\pm0.09)\%$ in the hydroxy group of methanol, which is a factor of $\sim$7.5 lower than the D/H ratio in its methyl group. Further investigations are needed in order to understand the synthesis of deuterated complex organic molecules.


\begin{acknowledgements}
We acknowledge support by the Deutsche Forschungsgemeinschaft via the collaborative 
research center SFB~956 (project ID 184018867) project B3 and SFB~1601 (project ID 500700252) projects A4 and Inf as well as the Ger{\"a}tezentrum SCHL~341/15-1 (``Cologne Center for Terahertz Spectroscopy''). 
The research in Kharkiv and Braunschweig was carried out under support of the Volkswagen foundation. 
The assistance of the Science and Technology Center in the Ukraine is acknowledged (STCU partner project P756). 
J.K.J. is supported by the Independent Research Fund Denmark (grant number 0135-00123B). 
R.M.L. received support from the Natural Sciences and Engineering Research Council of Canada. 
M.N.D. acknowledges the Holcim Foundation Stipend, the Swiss National Science Foundation (SNSF) Ambizione grant number 180079, the Center for Space and Habitability (CSH) Fellowship, and the IAU Gruber Foundation Fellowship. V.V.I. acknowledges financial support from Deutsche Forschungsgemeinschaft (grant number BA2176/9-1). E. Alekseev gratefully acknowledges financial support from Centre National de la Recherche Scientifique (CNRS, France) and from Universit{\'e} de Lille (France).  Our research benefited from NASA's Astrophysics Data System (ADS). 
This paper makes use of the following ALMA data: ADS/JAO.ALMA \# 2013.1.00278.S. 
ALMA is a partnership of ESO (representing its member states), NSF (USA) and NINS (Japan), together with 
NRC (Canada), MOST and ASIAA (Taiwan), and KASI (Republic of Korea), in cooperation with the Republic of Chile. 
The Joint ALMA Observatory is operated by ESO, AUI/NRAO and NAOJ.
\end{acknowledgements}


\bibliographystyle{aa} 
\bibliography{CH3OD} 

\onecolumn

\begin{appendix}
\section{Parameters of the RAM Hamiltonian for the CH$_3$OD molecule}


\begin{longtable}{lllr}
\caption{\label{tbl:ParametersTable} Fitted parameters of the RAM Hamiltonian for the CH$_3$OD molecule.}\\

\hline\hline $n_{tr}$\textit{$^a$} & Par.\textit{$^{b}$} & Operator\textit{$^c$} & Value\textit{$^{d,e}$} \\
\hline
\endfirsthead
\caption{continued.}\\
\hline\hline $n_{tr}$\textit{$^a$} & Par.\textit{$^{b}$} & Operator\textit{$^c$} & Value\textit{$^{d,e}$} \\
\hline
\endhead
\hline

 $  2_{ 2, 0}$ &    $(1/2)V_3$     &   $(1-\cos 3\alpha)$                                        &  $           183.171569(21)$ \\
 $  2_{ 2, 0}$ &    $F$            &   $p_\alpha^2$                                              &  $          17.42797209(17)$ \\
 $  2_{ 1, 1}$ &    $\rho$         &   $P_ap_\alpha$                                             &  $          0.6993446726(21)$ \\
 $  2_{ 0, 2}$ &    $A_{RAM}$            &   $P_a^2$                                                   &  $             3.675099(14) $ \\
 $  2_{ 0, 2}$ &    $B_{RAM}$            &   $P_b^2$                                                   &  $              0.783150(12) $ \\
 $  2_{ 0, 2}$ &    $C_{RAM}$            &   $P_c^2$                                                   &  $              0.733527(12) $ \\
 $  2_{ 0, 2}$ &    $2D_{ab}$      &   $(1/2)\{P_a{,}P_b\}$                                      &  $            0.055955652(61) $ \\
 $  4_{ 4, 0}$ &    $(1/2)V_6$     &   $(1-\cos 6\alpha)$                                        &  $             -0.807349(95)$ \\
 $  4_{ 4, 0}$ &    $F_m$          &   $p_\alpha^4$                                              &  $            -0.2945328(17) \times 10^{ -2}$ \\
 $  4_{ 3, 1}$ &    $\rho_m$       &   $P_ap_\alpha^3$                                           &  $           -0.11189627(44) \times 10^{ -1}$ \\
 $  4_{ 2, 2}$ &    $V_{3J}$       &   $P^2(1-\cos 3\alpha)$                                     &  $            -0.2211486(94) \times 10^{ -2}$ \\
 $  4_{ 2, 2}$ &    $V_{3K}$       &   $P_a^2(1-\cos 3\alpha)$                                   &  $              0.125539(11) \times 10^{ -1}$ \\
 $  4_{ 2, 2}$ &    $V_{3bc}$      &   $(P_b^2-P_c^2)(1-\cos 3\alpha)$                           &  $             -0.137249(22) \times 10^{ -3}$ \\
 $  4_{ 2, 2}$ &    $V_{3ab}$      &   $(1/2)\{P_a{,}P_b\}(1-\cos 3\alpha)$                      &  $            0.15619469(62) \times 10^{ -1}$ \\
 $  4_{ 2, 2}$ &    $F_J$          &   $P^2p_\alpha^2$                                           &  $            -0.8417192(26) \times 10^{ -4}$ \\
 $  4_{ 2, 2}$ &    $F_K$          &   $P_a^2p_\alpha^2$                                         &  $           -0.16465226(55) \times 10^{ -1}$ \\
 $  4_{ 2, 2}$ &    $F_{bc}$       &   $(P_b^2-P_c^2)p_\alpha^2$                                 &  $            -0.8781804(46) \times 10^{ -4}$ \\
 $  4_{ 2, 2}$ &    $F_{ab}$       &   $(1/2)\{P_a{,}P_b\}p_\alpha^2$                            &  $              0.123059(61) \times 10^{ -3}$ \\
 $  4_{ 2, 2}$ &    $D_{3ac}$      &   $(1/2)\{P_a{,}P_c\}\sin 3\alpha$                          &  $              0.287840(14) \times 10^{ -1}$ \\
 $  4_{ 1, 3}$ &    $\rho_J$       &   $P^2P_ap_\alpha$                                          &  $           -0.12144162(39) \times 10^{ -3}$ \\
 $  4_{ 1, 3}$ &    $\rho_K$       &   $P_a^3p_\alpha$                                           &  $           -0.10794357(36) \times 10^{ -1}$ \\
 $  4_{ 1, 3}$ &    $\rho_{bc}$    &   $(1/2)\{P_a{,}(P_b^2-P_c^2)\}p_\alpha$                    &  $            -0.1589380(32) \times 10^{ -3}$ \\
 $  4_{ 1, 3}$ &    $\rho_{ab}$    &   $(1/2)\{P_a^2{,}P_b\}p_\alpha$                            &  $              0.115410(57) \times 10^{ -3}$ \\
 $  4_{ 0, 4}$ &    $-\Delta_J$    &   $P^4$                                                     &  $             -0.144728(12) \times 10^{ -5}$ \\
 $  4_{ 0, 4}$ &    $-\Delta_{JK}$ &   $P^2P_a^2$                                                &  $              -0.46863(83) \times 10^{ -4}$ \\
 $  4_{ 0, 4}$ &    $-\Delta_K$    &   $P_a^4$                                                   &  $           -0.26562002(84) \times 10^{ -2}$ \\
 $  4_{ 0, 4}$ &    $-2\delta_J$   &   $P^2(P_b^2-P_c^2)$                                        &  $            -0.1984984(31) \times 10^{ -6}$ \\
 $  4_{ 0, 4}$ &    $-2\delta_K$   &   $(1/2)\{P_a^2{,}(P_b^2-P_c^2)\}$                          &  $             -0.726533(26) \times 10^{ -4}$ \\
 $  4_{ 0, 4}$ &    $D_{abJ}$      &   $(1/2)P^2\{P_a{,}P_b\}$                                   &  $              -0.69723(14) \times 10^{ -6}$ \\
 $  6_{ 6, 0}$ &    $(1/2)V_9$     &   $(1-\cos 9\alpha)$                                        &  $                0.1588(23) \times 10^{ -1}$ \\
 $  6_{ 6, 0}$ &    $F_{mm}$       &   $p_\alpha^6$                                              &  $               0.22823(16) \times 10^{ -5}$ \\
 $  6_{ 5, 1}$ &    $\rho_{mm}$    &   $P_ap_\alpha^5$                                           &  $              0.150802(69) \times 10^{ -4}$ \\
 $  6_{ 4, 2}$ &    $V_{6J}$       &   $P^2(1-\cos 6\alpha)$                                     &  $               -0.6128(78) \times 10^{ -4}$ \\
 $  6_{ 4, 2}$ &    $V_{6K}$       &   $P_a^2(1-\cos 6\alpha)$                                   &  $                0.1056(50) \times 10^{ -3}$ \\
 $  6_{ 4, 2}$ &    $V_{6bc}$      &   $(P_b^2-P_c^2)(1-\cos 6\alpha)$                           &  $              -0.28947(72) \times 10^{ -4}$ \\
 $  6_{ 4, 2}$ &    $V_{6ab}$      &   $(1/2)\{P_a{,}P_b\}(1-\cos 6\alpha)$                      &  $               -0.2335(10) \times 10^{ -4}$ \\
 $  6_{ 4, 2}$ &    $F_{mJ}$       &   $P^2p_\alpha^4$                                           &  $               0.25716(24) \times 10^{ -7}$ \\
 $  6_{ 4, 2}$ &    $F_{mK}$       &   $P_a^2p_\alpha^4$                                         &  $               0.40075(13) \times 10^{ -4}$ \\
 $  6_{ 4, 2}$ &    $D_{6ac}$      &   $(1/2)\{P_a{,}P_c\}\sin 6\alpha$                          &  $               0.10499(12) \times 10^{ -3}$ \\
 $  6_{ 3, 3}$ &    $\rho_{mJ}$    &   $P^2P_ap_\alpha^3$                                        &  $               0.89234(68) \times 10^{ -7}$ \\
 $  6_{ 3, 3}$ &    $\rho_{mK}$    &   $P_a^3p_\alpha^3$                                         &  $               0.55525(14) \times 10^{ -4}$ \\
 $  6_{ 3, 3}$ &    $\rho_{3bc}$   &   $(1/2)\{P_a{,}P_b{,}P_c{,}p_\alpha{,}\sin 3\alpha\}$      &  $               0.17326(13) \times 10^{ -5}$ \\
 $  6_{ 2, 4}$ &    $V_{3JJ}$      &   $P^4(1-\cos 3\alpha)$                                     &  $               0.13288(18) \times 10^{ -7}$ \\
 $  6_{ 2, 4}$ &    $V_{3JK}$      &   $P^2P_a^2(1-\cos 3\alpha)$                                &  $             -0.108198(83) \times 10^{ -5}$ \\
 $  6_{ 2, 4}$ &    $V_{3KK}$      &   $P_a^4(1-\cos 3\alpha)$                                   &  $               0.12335(14) \times 10^{ -5}$ \\
 $  6_{ 2, 4}$ &    $V_{3bcJ}$     &   $P^2(P_b^2-P_c^2)(1-\cos 3\alpha)$                        &  $               0.54564(17) \times 10^{ -8}$ \\
 $  6_{ 2, 4}$ &    $V_{3bcK}$     &   $(1/2)\{P_a^2{,}(P_b^2-P_c^2)\}(1-\cos 3\alpha)$          &  $               -0.6019(59) \times 10^{ -7}$ \\
 $  6_{ 2, 4}$ &    $V_{3b2c2}$    &   $(1/2)\{P_b^2{,}P_c^2\}\cos 3\alpha$                      &  $               0.36950(15) \times 10^{ -7}$ \\
 $  6_{ 2, 4}$ &    $V_{3abJ}$     &   $(1/2)P^2\{P_a{,}P_b\}(1-\cos 3\alpha)$                   &  $              -0.34986(11) \times 10^{ -6}$ \\
 $  6_{ 2, 4}$ &    $V_{3abK}$     &   $(1/2)\{P_a^3{,}P_b\}(1-\cos 3\alpha)$                    &  $              -0.16693(13) \times 10^{ -5}$ \\
 $  6_{ 2, 4}$ &    $V_{3abc2}$    &   $(1/2)\{P_a{,}P_b{,}P_c^2\}\cos 3\alpha$                  &  $              -0.41168(30) \times 10^{ -6}$ \\
 $  6_{ 2, 4}$ &    $F_{JJ}$       &   $P^4p_\alpha^2$                                           &  $               0.52398(53) \times 10^{ -9}$ \\
 $  6_{ 2, 4}$ &    $F_{JK}$       &   $P^2P_a^2p_\alpha^2$                                      &  $              0.120708(76) \times 10^{ -6}$ \\
 $  6_{ 2, 4}$ &    $F_{KK}$       &   $P_a^4p_\alpha^2$                                         &  $              0.426507(90) \times 10^{ -4}$ \\
 $  6_{ 2, 4}$ &    $F_{bcJ}$      &   $P^2(P_b^2-P_c^2)p_\alpha^2$                              &  $                 0.747(35) \times 10^{ -9}$ \\
 $  6_{ 2, 4}$ &    $F_{abJ}$      &   $(1/2)P^2\{P_a{,}P_b\}p_\alpha^2$                         &  $               -0.1940(17) \times 10^{ -8}$ \\
 $  6_{ 2, 4}$ &    $D_{3acJ}$     &   $(1/2)P^2\{P_a{,}P_c\}\sin 3\alpha$                       &  $              -0.21496(33) \times 10^{ -6}$ \\
 $  6_{ 2, 4}$ &    $D_{3acK}$     &   $(1/2)\{P_a^3{,}P_c\}\sin 3\alpha$                        &  $              -0.27643(22) \times 10^{ -5}$ \\
 $  6_{ 2, 4}$ &    $D_{3bcJ}$     &   $(1/2)P^2\{P_b{,}P_c\}\sin 3\alpha$                       &  $               -0.1440(78) \times 10^{ -7}$ \\
 $  6_{ 2, 4}$ &    $D_{3acb2}$    &   $(1/2)\{P_a{,}P_b^2{,}P_c\}\sin 3\alpha$                  &  $              -0.63621(23) \times 10^{ -6}$ \\
 $  6_{ 2, 4}$ &    $D_{3bcbc}$    &   $(1/2)(\{P_b^3{,}P_c\}-\{P_b{,}P_c^3\})\sin 3\alpha$      &  $              -0.15526(13) \times 10^{ -7}$ \\
 $  6_{ 1, 5}$ &    $\rho_{JJ}$    &   $P^4P_ap_\alpha$                                          &  $               0.76489(73) \times 10^{ -9}$ \\
 $  6_{ 1, 5}$ &    $\rho_{JK}$    &   $P^2P_a^3p_\alpha$                                        &  $               0.72735(40) \times 10^{ -7}$ \\
 $  6_{ 1, 5}$ &    $\rho_{KK}$    &   $P_a^5p_\alpha$                                           &  $              0.173086(31) \times 10^{ -4}$ \\
 $  6_{ 1, 5}$ &    $\rho_{bcJ}$   &   $(1/2)P^2\{P_a{,}(P_b^2-P_c^2)\}p_\alpha$                 &  $                0.1779(33) \times 10^{ -8}$ \\
 $  6_{ 0, 6}$ &    $\Phi_J$       &   $P^6$                                                     &  $                -0.839(30) \times 10^{-13}$ \\
 $  6_{ 0, 6}$ &    $\Phi_{JK}$    &   $P^4P_a^2$                                                &  $               0.30696(29) \times 10^{ -9}$ \\
 $  6_{ 0, 6}$ &    $\Phi_{KJ}$    &   $P^2P_a^4$                                                &  $               0.17819(12) \times 10^{ -7}$ \\
 $  6_{ 0, 6}$ &    $\Phi_K$       &   $P_a^6$                                                   &  $              0.290869(45) \times 10^{ -5}$ \\
 $  6_{ 0, 6}$ &    $2\phi_J$      &   $P^4(P_b^2-P_c^2)$                                        &  $                0.6901(12) \times 10^{-12}$ \\
 $  6_{ 0, 6}$ &    $2\phi_{JK}$   &   $(1/2)P^2\{P_a^2{,}(P_b^2-P_c^2)\}$                       &  $              0.102073(94) \times 10^{ -8}$ \\
 $  6_{ 0, 6}$ &    $2\phi_K$      &   $(1/2)\{P_a^4{,}(P_b^2-P_c^2)\}$                          &  $                0.2973(18) \times 10^{ -8}$ \\
 $  6_{ 0, 6}$ &    $D_{b2c2bc}$   &   $(1/2)(\{P_b^4{,}P_c^2\}-\{P_b^2{,}P_c^4\})$              &  $              -0.32957(53) \times 10^{-11}$ \\
 $  6_{ 0, 6}$ &    $D_{abJK}$     &   $(1/2)P^2\{P_a^3{,}P_b\}$                                 &  $                0.1662(15) \times 10^{ -8}$ \\
 $  6_{ 0, 6}$ &    $D_{abc4}$     &   $(1/2)\{P_a{,}P_b{,}P_c^4\}$                              &  $                0.1892(15) \times 10^{-10}$ \\
 $  8_{ 8, 0}$ &    $F_{mmm}$      &   $p_\alpha^8$                                              &  $                0.2592(24) \times 10^{ -8}$ \\
 $  8_{ 6, 2}$ &    $V_{9J}$       &   $P^2(1-\cos 9\alpha)$                                     &  $                0.3021(47) \times 10^{ -3}$ \\
 $  8_{ 6, 2}$ &    $V_{9K}$       &   $P_a^2(1-\cos 9\alpha)$                                   &  $                -0.657(12) \times 10^{ -3}$ \\
 $  8_{ 6, 2}$ &    $F_{mmK}$      &   $P_a^2p_\alpha^6$                                         &  $               -0.5042(33) \times 10^{ -7}$ \\
 $  8_{ 6, 2}$ &    $D_{9bc}$      &   $(1/2)\{P_b{,}P_c\}\sin 9\alpha$                          &  $               0.78071(91) \times 10^{ -4}$ \\
 $  8_{ 5, 3}$ &    $\rho_{mmK}$   &   $P_a^3p_\alpha^5$                                         &  $              -0.16973(94) \times 10^{ -6}$ \\
 $  8_{ 5, 3}$ &    $\rho_{3bcm}$  &   $(1/2)\{P_a{,}P_b{,}P_c{,}p_\alpha^3{,}\sin 3\alpha\}$    &  $                -0.770(13) \times 10^{ -8}$ \\
 $  8_{ 4, 4}$ &    $V_{6JJ}$      &   $P^4(1-\cos 6\alpha)$                                     &  $                0.2170(71) \times 10^{ -8}$ \\
 $  8_{ 4, 4}$ &    $V_{6JK}$      &   $P^2P_a^2(1-\cos 6\alpha)$                                &  $               -0.5615(58) \times 10^{ -6}$ \\
 $  8_{ 4, 4}$ &    $V_{6KK}$      &   $P_a^4(1-\cos 6\alpha)$                                   &  $                0.1808(27) \times 10^{ -6}$ \\
 $  8_{ 4, 4}$ &    $V_{6bcJ}$     &   $P^2(P_b^2-P_c^2)(1-\cos 6\alpha)$                        &  $               0.14457(44) \times 10^{ -8}$ \\
 $  8_{ 4, 4}$ &    $V_{6bcK}$     &   $(1/2)\{P_a^2{,}(P_b^2-P_c^2)\}(1-\cos 6\alpha)$          &  $                0.4650(41) \times 10^{ -7}$ \\
 $  8_{ 4, 4}$ &    $F_{mKK}$      &   $P_a^4p_\alpha^4$                                         &  $               -0.2700(13) \times 10^{ -6}$ \\
 $  8_{ 4, 4}$ &    $D_{6bcJ}$     &   $(1/2)P^2\{P_b{,}P_c\}\sin 6\alpha$                       &  $                -0.461(16) \times 10^{ -9}$ \\
 $  8_{ 4, 4}$ &    $D_{6bcbc}$    &   $(1/2)(\{P_b{,}P_c^3\}-\{P_b^3{,}P_c\})\sin 6\alpha$      &  $                0.2259(44) \times 10^{ -8}$ \\
 $  8_{ 4, 4}$ &    $D_{3acb2m}$   &   $(1/2)\{P_a{,}P_b^2{,}P_c{,}p_\alpha^2{,}\sin 3\alpha\}$  &  $                 0.859(12) \times 10^{ -9}$ \\
 $  8_{ 3, 5}$ &    $\rho_{mKK}$   &   $P_a^5p_\alpha^3$                                         &  $               -0.2464(11) \times 10^{ -6}$ \\
 $  8_{ 3, 5}$ &    $\rho_{3bcK}$  &   $(1/2)\{P_a^3{,}P_b{,}P_c{,}p_\alpha{,}\sin 3\alpha\}$    &  $                0.4437(26) \times 10^{ -7}$ \\
 $  8_{ 2, 6}$ &    $V_{3JJJ}$     &   $P^6(1-\cos 3\alpha)$                                     &  $                -0.576(19) \times 10^{-13}$ \\
 $  8_{ 2, 6}$ &    $V_{3KKK}$     &   $P_a^6(1-\cos 3\alpha)$                                   &  $               -0.3463(78) \times 10^{ -9}$ \\
 $  8_{ 2, 6}$ &    $V_{3bcKK}$    &   $(1/2)\{P_a^4{,}(P_b^2-P_c^2)\}(1-\cos 3\alpha)$          &  $               -0.2770(18) \times 10^{ -8}$ \\
 $  8_{ 2, 6}$ &    $V_{3b2c2bc}$  &   $(1/2)(\{P_b^4{,}P_c^2\}-\{P_b^2{,}P_c^4\})\cos 3\alpha$  &  $                0.7702(54) \times 10^{-12}$ \\
 $  8_{ 2, 6}$ &    $V_{3abJJ}$    &   $(1/2)P^4\{P_a{,}P_b\}(1-\cos 3\alpha)$                   &  $                0.3300(70) \times 10^{-11}$ \\
 $  8_{ 2, 6}$ &    $V_{3abc4}$    &   $(1/2)\{P_a{,}P_b{,}P_c^4\}\cos 3\alpha$                  &  $                0.1252(17) \times 10^{-10}$ \\
 $  8_{ 2, 6}$ &    $F_{JJJ}$      &   $P^6p_\alpha^2$                                           &  $                -0.363(21) \times 10^{-14}$ \\
 $  8_{ 2, 6}$ &    $F_{KKK}$      &   $P_a^6p_\alpha^2$                                         &  $              -0.13293(51) \times 10^{ -6}$ \\
 $  8_{ 2, 6}$ &    $D_{3acJJ}$    &   $(1/2)P^4\{P_a{,}P_c\}\sin 3\alpha$                       &  $               -0.2388(78) \times 10^{-11}$ \\
 $  8_{ 2, 6}$ &    $D_{3acKK}$    &   $(1/2)\{P_a^5{,}P_c\}\sin 3\alpha$                        &  $                0.3398(79) \times 10^{ -9}$ \\
 $  8_{ 2, 6}$ &    $D_{3bcJJ}$    &   $(1/2)P^4\{P_b{,}P_c\}\sin 3\alpha$                       &  $                0.4091(84) \times 10^{-12}$ \\
 $  8_{ 2, 6}$ &    $D_{3bcKK}$    &   $(1/2)\{P_a^4{,}P_b{,}P_c\}\sin 3\alpha$                  &  $                0.2557(16) \times 10^{ -7}$ \\
 $  8_{ 2, 6}$ &    $D_{3acb2J}$   &   $(1/2)P^2\{P_a{,}P_b^2{,}P_c\}\sin 3\alpha$               &  $                0.1579(16) \times 10^{-10}$ \\
 $  8_{ 2, 6}$ &    $D_{3acb2K}$   &   $(1/2)\{P_a^3{,}P_b^2{,}P_c\}\sin 3\alpha$                &  $               -0.5322(48) \times 10^{ -9}$ \\
 $  8_{ 2, 6}$ &    $D_{3bcbcJ}$   &   $(1/2)P^2(\{P_b^3{,}P_c\}-\{P_b{,}P_c^3\})\sin 3\alpha$   &  $                0.3740(28) \times 10^{-12}$ \\
 $  8_{ 1, 7}$ &    $\rho_{JJJ}$   &   $P^6P_ap_\alpha$                                          &  $                -0.710(29) \times 10^{-14}$ \\
 $  8_{ 1, 7}$ &    $\rho_{KKK}$   &   $P_a^7p_\alpha$                                           &  $               -0.3969(14) \times 10^{ -7}$ \\
 $  8_{ 0, 8}$ &    $L_J$          &   $P^8$                                                     &  $               -0.1082(61) \times 10^{-16}$ \\
 $  8_{ 0, 8}$ &    $L_{JJK}$      &   $P^6P_a^2$                                                &  $                -0.355(10) \times 10^{-14}$ \\
 $  8_{ 0, 8}$ &    $L_K$          &   $P_a^8$                                                   &  $               -0.5083(16) \times 10^{ -8}$ \\
 $  8_{ 0, 8}$ &    $2l_K$         &   $(1/2)\{P_a^6{,}(P_b^2-P_c^2)\}$                          &  $                -0.254(13) \times 10^{-12}$ \\
 $ 10_{ 8, 2}$ &    $V_{12J}$      &   $P^2(1-\cos 12\alpha)$                                    &  $                -0.992(16) \times 10^{ -3}$ \\
 $ 10_{ 8, 2}$ &    $V_{12bc}$     &   $(P_b^2-P_c^2)(1-\cos 12\alpha)$                          &  $               -0.8712(96) \times 10^{ -4}$ \\
 $ 10_{ 6, 4}$ &    $V_{9JJ}$      &   $P^4(1-\cos 9\alpha)$                                     &  $                -0.326(18) \times 10^{ -8}$ \\
 $ 10_{ 6, 4}$ &    $V_{9JK}$      &   $P^2P_a^2(1-\cos 9\alpha)$                                &  $                0.2026(32) \times 10^{ -5}$ \\
 $ 10_{ 6, 4}$ &    $V_{9b2c2}$    &   $(1/2)\{P_b^2{,}P_c^2\}\cos 9\alpha$                      &  $                 0.544(16) \times 10^{ -8}$ \\
 $ 10_{ 6, 4}$ &    $D_{9acK}$     &   $(1/2)\{P_a^3{,}P_c\}\sin 9\alpha$                        &  $               -0.5240(98) \times 10^{ -6}$ \\
 $ 10_{ 6, 4}$ &    $D_{6acmK}$    &   $(1/2)\{P_a^3{,}P_c{,}p_\alpha^2{,}\sin 6\alpha\}$        &  $                0.1717(33) \times 10^{ -7}$ \\
 $ 10_{ 4, 6}$ &    $V_{6JJJ}$     &   $P^6(1-\cos 6\alpha)$                                     &  $                -0.551(35) \times 10^{-13}$ \\
 $ 10_{ 4, 6}$ &    $V_{6JKK}$     &   $P^2P_a^4(1-\cos 6\alpha)$                                &  $                0.8775(30) \times 10^{ -9}$ \\
 $ 10_{ 4, 6}$ &    $V_{6KKK}$     &   $P_a^6(1-\cos 6\alpha)$                                   &  $                 0.214(14) \times 10^{ -9}$ \\
 $ 10_{ 4, 6}$ &    $V_{6bcJJ}$    &   $P^4(P_b^2-P_c^2)(1-\cos 6\alpha)$                        &  $                -0.808(19) \times 10^{-13}$ \\
 $ 10_{ 4, 6}$ &    $V_{6b2c2K}$   &   $(1/2)\{P_a^2{,}P_b^2{,}P_c^2\}\cos 6\alpha$              &  $                -0.580(19) \times 10^{-10}$ \\
 $ 10_{ 4, 6}$ &    $D_{6acKK}$    &   $(1/2)\{P_a^5{,}P_c\}\sin 6\alpha$                        &  $                -0.842(15) \times 10^{ -8}$ \\
 $ 10_{ 3, 7}$ &    $\rho_{3bcKK}$ &   $(1/2)\{P_a^5{,}P_b{,}P_c{,}p_\alpha{,}\sin 3\alpha\}$    &  $                0.1553(41) \times 10^{-11}$ \\
 $ 10_{ 2, 8}$ &    $V_{3JKKK}$    &   $P^2P_a^6(1-\cos 3\alpha)$                                &  $                 0.548(14) \times 10^{-13}$ \\
 $ 10_{ 2, 8}$ &    $D_{3b3c3J}$   &   $(1/2)P^2\{P_b^3{,}P_c^3\}\sin 3\alpha$                   &  $                -0.977(46) \times 10^{-16}$ \\
 $ 12_{ 8, 4}$ &    $V_{12JK}$     &   $P^2P_a^2(1-\cos 12\alpha)$                               &  $                -0.408(11) \times 10^{ -5}$ \\
 $ 12_{ 6, 6}$ &    $V_{9JKK}$     &   $P^2P_a^4(1-\cos 9\alpha)$                                &  $              -0.23634(86) \times 10^{ -8}$ \\
 $ 12_{ 6, 6}$ &    $V_{9b2c2K}$   &   $(1/2)\{P_a^2{,}P_b^2{,}P_c^2\}\cos 9\alpha$              &  $                 0.639(42) \times 10^{-10}$ \\
 $ 12_{ 4, 8}$ &    $D_{6bcJJK}$   &   $(1/2)P^4\{P_a^2{,}P_b{,}P_c\}\sin 6\alpha$               &  $                 0.492(18) \times 10^{-14}$ \\

\hline
\hline

\end{longtable}

\tablefoot{$^{a}$ \textit{n=t+r}, where \textit{n} is the total order of the operator, \textit{t} is the order 
  of the torsional part, and \textit{r} is the order of the rotational part, respectively. The ordering scheme 
  of \citet{Nakagawa:1987} is used. $^{b}$ The parameter nomenclature is based on the subscript 
  procedure of \citet{CH3OH_rot_2008}. $^{c}$ $\lbrace A,B,C,D,E \rbrace = ABCDE+EDCBA$.  $\lbrace A,B,C,D \rbrace = ABCD+DCBA$. $\lbrace A,B,C \rbrace = ABC+CBA$. $\lbrace A,B \rbrace = AB+BA$. 
  The product of the operator in the third column of a given row and the parameter in the second column 
  of that row gives the term actually used in the torsion-rotation Hamiltonian of the program, except for 
  \textit{F}, $\rho$, and \textit{$A_{\rm RAM}$}, which occur in the Hamiltonian in the form 
  $F(p_\alpha + \rho P_a)^2 + A_{\rm RAM}P_a^2$.  $^{d}$ Values of the parameters are in units of reciprocal centimeters, except for $\rho$, 
  which is unitless. $^{e}$ Statistical uncertainties are given in parentheses as one standard uncertainty 
  in units of the last digits.}

\clearpage

\section{Corner plot of the second MCMC run of the CH$_{3}$OD computation for IRAS~16293-2422~B.}
\begin{figure}[hbt!]
	\includegraphics[width=0.5\hsize]{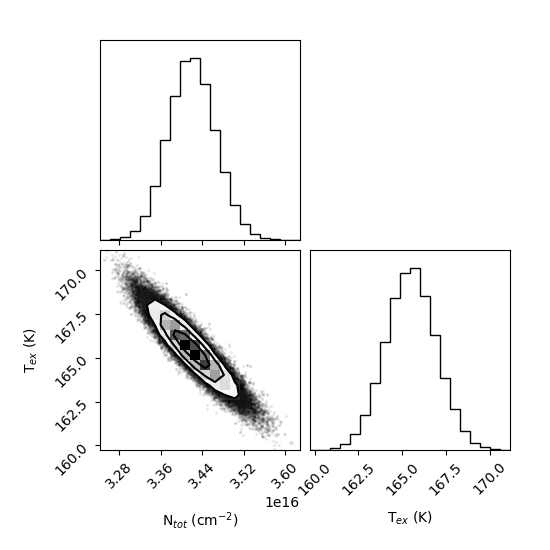}
	\caption{Corner plot of the second MCMC run of the CH$_{3}$OD computation for IRAS~16293-2422~B.}
	\label{fig:CH3OD_corner}
\end{figure}

\end{appendix}

\end{document}